# Evaluation of Deep Learning Models for LBBB Classification in ECG Signals

Beatriz Macas Ordóñez[1], Diego Vinicio Orellana Villavicencio[2], José Manuel Ferrández[3] and Paula Bonomini[1]

*Abstract*—This study explores different neural network architectures to evaluate their ability to extract spatial and temporal patterns from electrocardiographic (ECG) signals and classify them into three groups: healthy subjects, Left Bundle Branch Block (LBBB), and Strict Left Bundle Branch Block (sLBBB).

*Clinical Relevance*—Innovative technologies enable the selection of candidates for Cardiac Resynchronization Therapy (CRT) by optimizing the classification of subjects with Left Bundle Branch Block (LBBB).

## I. INTRODUCTION

Left bundle branch block (LBBB) may predict a better response to cardiac resynchronization therapy (CRT). This study evaluates different deep learning architectures for the classification of electrocardiographic (ECG) signals from healthy subjects, patients with Left Bundle Branch Block (LBBB), and Strict LBBB (sLBBB). By applying wavelet denoising, Principal Component Analysis (PCA), and convolutional feature extraction, the models—CNN, GRU, LSTM, Attention, Bi-GRU, and Bi-LSTM—are trained and assessed individually.

## II. MATERIALS AND METHODS

Two datasets were analyzed in this study. The first dataset consisted of normal electrocardiographic recordings, while the second included a cohort of heart failure patients from the MADIT-CRT clinical trial, available through the THEW project at the University of Rochester. The database comprised three ECG categories: healthy (n=299), LBBB (n=192), and sLBBB (n=301), all recorded with 12 leads.

To improve signal quality, Wavelet Transform was applied to reduce noise, followed by principal component analysis (PCA) to reduce dimensionality. The signals were then normalized and reformatted to preserve both spatial and temporal characteristics, ensuring compatibility with deep learning models.

Six neural network architectures—CNN, GRU, LSTM, attention-based mechanisms, Bi-GRU, and Bi-LSTM—were trained independently within a standardized framework. Each model was optimized using AdamW and monitored through early stopping, adaptive learning rate adjustments, and model checkpointing. Performance evaluation was conducted using accuracy metrics, confusion matrices, and classification reports, allowing a comparative assessment of their effectiveness in distinguishing between the three ECG categories.

[1]*Instituto Argentino de Matematica "Alberto P. Calderon" (IAM), CONICET, Buenos Aires, Argentina.* bmacas@fi.uba.ar

[2]*Universidad Nacional de Loja (UNL), Ecuador*

[3]*Depto. de Electrónica, Tecnología de Computadoras y Proyectos, Universidad Politecnica de Cartagena, Cartagena, Spain*

## III. RESULTS

The evaluation of deep learning architectures for ECG classification revealed that the Bi-LSTM model achieved the highest accuracy (91.52%), outperforming CNN, GRU, LSTM, Attention, and Bi-GRU.

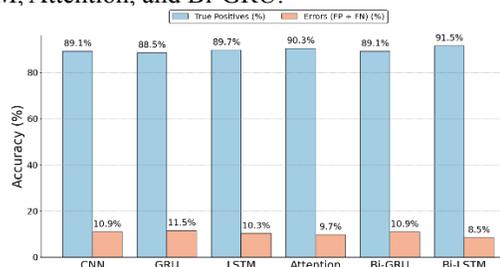

1. Accuracy comparison (vertical axis) of different deep learning models (horizontal axis) for ECG classification, showing True Positives and Errors.

## IV. DISCUSSION AND CONCLUSIONS

The results show that the Bi-LSTM model achieved the highest accuracy (91.52%), indicating that bidirectional architectures are effective at capturing temporal dependencies in ECG signals. In contrast, the CNN model (89.09%) performed well but struggled with healthy classifications, likely due to its focus on spatial rather than temporal features. The GRU and Bi-GRU models (88.48% and 89.09%) showed comparable performance but had similar issues with healthy ECG patterns. The Attention-based model (90.30%) provided more balanced results, suggesting the effectiveness of self-attention mechanisms in refining ECG feature selection. While non-Bi-LSTM models showed strengths, they struggled with recall in at least one class, highlighting the importance of bidirectional networks for robust ECG classification. To compare deep learning and traditional machine learning approaches, we also implemented a Support Vector Machine (SVM). The SVM achieved an accuracy of 87%, highlighting that deep learning architectures specifically designed for time series data demonstrate superior potential for processing ECG signals.

## ACKNOWLEDGMENT

This research was funded by Grant 26-DI-FEIRNNR- 2023 from the Universidad Nacional de Loja (Ecuador) and partially supported by "*Programa Iberoamericano de Ciencia y Tecnología para el Desarrollo (CYTED)*" through Red 225RT0169.